\documentclass[preprint,12pt]{elsarticle}

\usepackage{float}

\usepackage{kpfonts}
\usepackage[margin = 1in]{geometry}
\usepackage{booktabs}
\usepackage{changepage}

\usepackage{graphicx} 
\usepackage{xcolor}
\usepackage[colorlinks = true,
linkcolor = blue,
urlcolor  = blue,
citecolor = blue,
anchorcolor = blue]{hyperref}
\usepackage{tikz}
\usepackage{caption}

\usepackage{amssymb}
\usepackage{amsmath}

\usepackage{lineno}

\newcommand{\mat}[1]{\mathbf{#1}}
\newcommand{\vecb}[1]{\mathbf{#1}}

\journal{Theoretical Biology}

\begin{document}
	
	\begin{frontmatter}
		
		
		
		\title{Life Finds a Way: Emergence of Cooperative Structures in Adaptive 
		Threshold Networks}
		
		\author[label1,label2]{Sean P Maley}
		\author[label2]{Carlos Gershenson} 
		\author[label3]{Stuart A Kauffman} 
		\affiliation[label1]{organization={SUNY Finger Lakes CC},
			addressline={3325 Marvin Sands Dr},
			city={Canandaigua},
			postcode={14424},
			state={NY},
			country={USA}}
		
		\affiliation[label2]{organization={SUNY Binghamton},
			addressline={4400 Vestal Pkwy E},
			city={Binghamton},
			postcode={13902},
			state={NY},
			country={USA}}
		
		\affiliation[label3]{organization={Institute for Systems Biology},
			addressline={401 Terry Ave N},
			city={Seattle},
			postcode={98109},
			state={WA},
			country={USA}}

		
		\begin{abstract}
			How new levels of organization arise is a long-standing question. The
			emergence of higher-level organization can appear unlikely, since
			cooperation must prevail over competition. One well-studied example is the
			autocatalytic set, a collection of molecules in which the formation of each
			member is catalyzed by another member of the set, and which is often
			considered a prerequisite for the evolution of life.
			
			We present a random threshold-directed network model that integrates
			node-specific traits with dynamic edge formation and node removal,
			simulating arbitrary levels of cooperation and competition. Intrinsic node
			values determine directed links through threshold rules, generating a
			multi-digraph with signed edges (reflecting support and antagonism, labeled
			``help'' and ``harm'') that yields two parallel yet interdependent threshold
			graphs. Incorporating temporal growth and node turnover allows exploration
			of the evolution, adaptation, and potential collapse of communities.
			
			We find that a strongly connected core assembles and persists even when
			antagonistic interactions predominate. Members are culled once incoming harm
			outweighs incoming help, so those that remain are helped more than they are
			harmed, and a quantitative increase in the number of elements produces a
			qualitative transition. As the harm-to-help ratio
			$\rho$ rises, late-time population growth and mean connectivity both decline
			steadily, growth reaching zero and connectivity its floor near
			$\rho_c \approx 0.6$, marking a shift from sustained growth to a bounded
			regime. Raising the binding chance
			moves the onset of a system-spanning strongly connected component to
			progressively smaller populations without changing its eventual extent,
			indicating that interaction opportunity governs when collective
			organization appears rather than whether it appears.
			
			These results extend classical random threshold and Erdős-Rényi random
			graph models, and bear on how microbial communities and other adaptive
			systems assemble Collective Affordance Sets under persistent antagonism.
		\end{abstract}
		
		\begin{graphicalabstract}
			\begin{center}
				\includegraphics[width=1\linewidth]{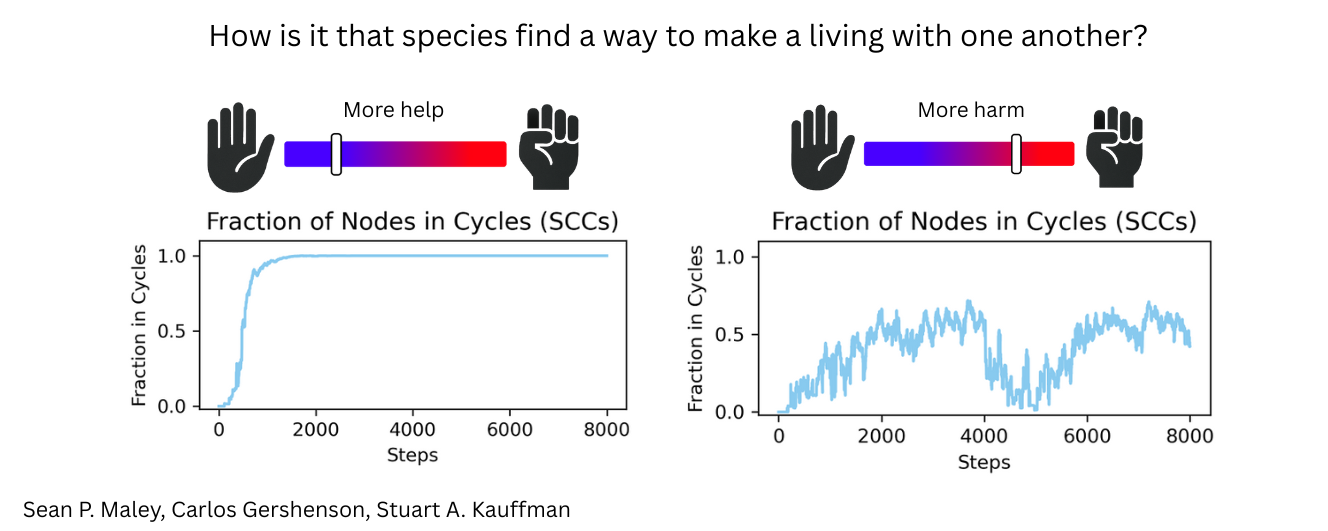}
			\end{center}
			
		\end{graphicalabstract}
		
		
		\begin{highlights}

			\item ``Order for free'': a mutually enabling SCC self-assembles despite
			antagonism.
			\item Quantity becomes quality at a threshold: a giant SCC takes shape.
			\item At high harm probabilities the community does not die: growth halts,
			core rebuilds.

		\end{highlights}

		
		
		\begin{keyword}
			adaptive threshold networks \sep cooperation under antagonism \sep 
			strongly connected components \sep regime change \sep crossover  \sep collective 
			affordance sets \sep emergent organization \sep complex systems biology 
			\sep microbial communities
		\end{keyword}

		

	\end{frontmatter}

	\usetikzlibrary{arrows,decorations.pathmorphing,backgrounds,positioning,fit}
	\usetikzlibrary{positioning}
	
		
		
		\section{Introduction}

		Life drives --- and is driven by --- interactions.   Each organism brings a 
		unique and vast combination of causal properties into its environment, 
		continuously adapting and “tinkering” to establish beneficial interactions 
		or mitigate harmful ones. Despite evolutionary pressures and competition, we 
		regularly observe that complex, self-sustaining networks of mutual support 
		routinely emerge in living systems~\citep{nowak_2006_five}. Understanding 
		how 
		and why such networks reliably form is central to both theoretical biology 
		and complex systems science.
		
		Microbial communities illustrate this phenomenon particularly well. Despite
		(or, possibly due to) immense biochemical and ecological diversity, stable
		interdependent communities arise repeatedly. Even when initial interactions
		are predominantly antagonistic, microbes find ways to coexist or even
		benefit each other: both antagonistic and synergistic interactions are
		present in bacterial communities, including those sampled directly from
		soil, and it is the mixture of the two rather than the dominance of either
		that appears to maintain diversity and buffer the community against
		perturbation~\citep{mall2022limited}. Such communities can
		be viewed as "collective affordance
		sets"~\citep{kauffman2023third}: jury-rigged, self-organized structures
		that exploit combinations of causal properties among community members.
		
		Assembling such communities deliberately, by combining many microbial
		isolates and observing whether novel stable ecosystems emerge from the vast
		space of possible interaction patterns, is now an established experimental
		paradigm with codified cross-system
		practice~\citep{mehlferber2024cross,coker2022reproducible,madsen2025synthetic}.
		Our primary interest is in complex soil communities of microorganisms, though 
		the same pattern is visible wherever new elements enter a system and find 
		niches among the ones already there; we return to that reach in Section~\ref{sec:diverse}.
		
		Cooperative and interdependent structures in microbial 
		communities need not arise from explicitly altruistic strategies. A 
		prominent alternative framework is the Black Queen
		Hypothesis~\citep{morris2012black}, which
		explains the emergence of metabolic dependencies through adaptive gene loss
		and the retention of “leaky” functions by a subset of community members. In
		this view, organisms may reduce individual costs by abandoning costly 
		functions while relying on others to supply shared resources, producing 
		cooperation-like patterns without cooperative intent. The present work is 
		not intended to model gene loss or metabolic specialization directly. 
		Rather, it focuses on the dynamical consequences of structured help and harm 
		interactions at the network level, abstracting away from the evolutionary 
		pathways by which such interactions arise. As such, the framework developed 
		here should be viewed as complementary to Black Queen–type mechanisms, 
		addressing how interdependence (once present) shapes the stability and 
		organization of communities.
		
		How often microbes actually help rather than harm one another is, however,
		still an open empirical question, and the disagreement in the literature is
		substantial. Surveying bacterial strains isolated from a common
		environment, Foster and Bell found that the great majority of pairwise
		interactions were net negative, with no evidence that positive higher-order
		effects arise when more than two species are mixed, and concluded that the
		typical result of adaptation to other microbial species is a competitive
		rather than a cooperative phenotype~\citep{foster2012competition}.
		Measurements among soil isolates chosen for contrasting life-history
		strategies likewise returned interactions dominated by negative outcomes,
		with cross-feeding accounting only for the relatively rare positive
		ones~\citep{wang2022substrate}. Against this, a high-throughput screen
		measuring 180,408 interactions among twenty soil bacteria across forty
		carbon environments found positive interactions to be common rather than
		rare. Those positive interactions occur between strains that differ in their
		carbon consumption profiles, and the study attributes them to
		metabolite-mediated mechanisms such as enzyme secretion and overflow
		metabolism. Their sign structure is nonetheless asymmetric rather than
		reciprocal: primarily parasitisms, in which one strain is facilitated at the
		expense of the other, whose growth is
		inhibited~\citep{kehe2021positive}. Within that one study the numbers
		vary as much as the literature does: the fraction of pairs with at least one
		positive interaction ran from about a fifth to well over a half, depending
		only on the carbon source the pair was grown in. Synthetic communities shift
		the same way, tipping toward antagonism as nutrients run
		short~\citep{qiao2024nutrient}.

		This unsettled quantity is the central parameter of the model developed
		below. Rather than adopt any one reported value, we treat the harm-to-help
		ratio $\rho$ as a control parameter and sweep it across its full range, from
		purely cooperative to almost entirely antagonistic. If the balance shifts
		with the carbon source, there is no single value to adopt. The sweep also
		poses the question the disagreement makes interesting: whether cooperative
		cores can persist even at the antagonism-dominated end of the range that
		Foster and Bell's data suggest.

		What a mixture of beneficial and antagonistic interactions does to community
		stability is a long-established question. Random-matrix analyses of the
		community matrix set out which combinations of interaction strengths admit a
		stable equilibrium~\citep{allesina2012stability}, and the proportion of
		mutualistic links has been an explicit object of study for more than a decade:
		varying that proportion directly, Mougi and Kondoh report that a mixture of
		mutualism and antagonism can stabilize a community that neither type alone
		would~\citep{mougi2012diversity}, a result whose dependence on how interaction
		strengths are normalized was subsequently
		disentangled~\citep{suweis2014disentangling}; carried into microbiomes, the same
		approach argues that cooperation destabilizes a community while competition
		stabilizes it~\citep{coyte2015ecology}. The question we put to $\rho$ is that
		literature's question, but $\rho$ is not one minus a proportion of mutualistic
		links, and the difference is worth stating. Help and harm here are properties of
		a single directed edge, whereas an interaction type in the community-matrix
		sense is a property of both directions of a species pair taken together, so an
		interaction type is an outcome of this model rather than a parameter of it. The
		traits governing help and harm are drawn independently, as are those governing
		the two directions of a pair, so a realized edge is helpful with probability
		$1-\rho$, and a pair of reciprocal edges is mutualistic with probability
		$(1-\rho)^2$, exploitative with probability $2\rho(1-\rho)$ and competitive with
		probability $\rho^2$: about $16\%$, $48\%$ and $36\%$ respectively at the
		critical value reported below, $\rho_c \approx 0.6$. A help edge also costs the
		helper nothing, so it is nearer to facilitation than to a mutualism that carries
		a cost of cooperation.

		In this paper, we present an adaptive threshold network 
		model~\citep{reilly_scheinerman_zhang_2014} with simple 
		assumptions, designed to explore the conditions that foster stable, 
		cooperative structures. Rather than focus on specific evolutionary games or 
		particular ecological scenarios, our model investigates the broader space of 
		possible interaction patterns, identifying conditions under which large, 
		strongly connected communities (SCCs) reliably form and persist.  As Faust and 
		Raes observe, ``alternative community states can arise as a consequence of system dynamics 
		without being driven by environmental differences"~\citep{faust_raes_2012}, 
		which is what an environment-agnostic model is built to examine.
		
		Using a simple model, our results demonstrate that a strongly connected core, 
		whose members are helped more than they are harmed, robustly emerges even when 
		antagonistic interactions predominate, provided certain thresholds of 
		cooperation and species longevity are met. 
		Moreover, these emergent communities exhibit distinctive patterns, such as 
		stable configurations of key cooperative nodes and systematic turnover 
		influenced by antagonism. Thus, our findings offer insights not only into 
		the formation of biological ecosystems but also into analogous processes in 
		economics and social systems, highlighting universal principles underlying 
		collective organization and adaptive complexity.

\section{Model}

\subsection{Model definition (birth--binding--culling dynamics)}

We simulate an evolving signed directed multigraph
$G_t=(V_t,E_t)$ using NetworkX's \texttt{MultiDiGraph}, where vertices represent agents and directed edges represent directed interactions. Time is discrete, $t=1,2,\dots$, and each time-step consists of: (i) adding one new node, (ii) probabilistically forming directed help/harm edges between the newcomer and all incumbents, (iii) aging all nodes by one tick, and (iv) removing any nodes that become eligible for deletion.

\paragraph{Parameters.}
The dynamics are governed by six global parameters, defined in Table~\ref{tab:params}.

\begin{table}[htbp]
	\centering
	\caption{Global parameters governing the birth--binding--culling dynamics.}
	\label{tab:params}
	\begin{tabular}{@{}lp{0.55\linewidth}l@{}}
		\toprule
		\textbf{Symbol} & \textbf{Definition} & \textbf{Default} \\
		\midrule
		$\lambda>0$ & Poisson rate controlling heterogeneity in opportunities for
		pairwise interaction. & varies by simulation \\
		\addlinespace
		$p_{\max}\in(0,1]$ & Baseline upper bound used to convert opportunity into
		an effective binding threshold/probability. & varies by simulation \\
		\addlinespace
		$\rho\in[0,1]$ & Global help--harm mixture: the probability that a
		successful binding event is harmful (and $1-\rho$ that it is helpful). &
		swept across simulations \\
		\addlinespace
		$m_{\max}\in\mathbb{N}$ & Maximum number of parallel edges allowed per
		ordered pair and sign. & $6$ \\
		\addlinespace
		$L_S\in\mathbb{N}$ & Minimum age before a node becomes eligible for
		removal (a lifespan/juvenile protection period). & $100$ \\
		\addlinespace
		$\eta\in[0,1]$ & Removal threshold for the fraction of incoming harm. &
		$0.5$ \\
		\bottomrule
	\end{tabular}
\end{table}

\paragraph{Node traits (intrinsic capacities).}
When a new node $i$ enters, it is endowed with four independent traits, defined
in Table~\ref{tab:traits}, each drawn i.i.d. uniformly from $[0,1]$.

\begin{table}[htbp]
	\centering
	\caption{Intrinsic node traits, drawn i.i.d. $\sim\mathrm{Uniform}[0,1]$ at node
	creation.}
	\label{tab:traits}
	\begin{tabular}{@{}lp{0.7\linewidth}@{}}
		\toprule
		\textbf{Symbol} & \textbf{Definition} \\
		\midrule
		$h_{\text{in}}(i)$ & Receptivity to beneficial (help) interactions. \\
		\addlinespace
		$h_{\text{out}}(i)$ & Capacity to provide beneficial (help) interactions. \\
		\addlinespace
		$\gamma_{\text{in}}(i)$ & Vulnerability to detrimental (harm) interactions. \\
		\addlinespace
		$\gamma_{\text{out}}(i)$ & Capacity to inflict detrimental (harm) interactions. \\
		\bottomrule
	\end{tabular}
\end{table}

We treat help and harm as \emph{independent dimensions} rather than opposite ends of a single axis. Biologically, an organism can plausibly both support mutualists and inhibit competitors via distinct pathways, so the ability to help need not constrain the ability to harm (and vice versa).

\paragraph{Pair-specific binding opportunity and effective threshold.}
When node $i$ enters at time $t$, it ``tests'' every incumbent node $j\in V_{t-1}$ in \emph{both} directions $(i\to j)$ and $(j\to i)$. For each ordered pair $(u,v)\in\{(i,j),(j,i)\}$ we draw
\[
k_{uv}\sim\mathrm{Poisson}(\lambda),
\qquad
P_{uv}=\min\bigl(1,\;k_{uv}\,p_{\max}\bigr).
\]
Here $k_{uv}$ represents the (random) number of opportunities for $u$ to interact with $v$ during this time-step, and $P_{uv}$ converts that opportunity into an effective binding scale capped at $1$.

\paragraph{Edge formation (help vs harm, with multiplicity).}
For each ordered pair $(u,v)$ involving the newcomer and an incumbent, we perform up 
to $m_{\max}$ binding attempts. Each attempt proceeds as follows:
\begin{enumerate}
	\item Draw the interaction type: harm with probability $\rho$, help with probability $1-\rho$.
	\item Conditional on the type, form a directed edge if the relevant trait mismatch is within the effective threshold:
	\[
	\text{help:}\quad |h_{\text{out}}(u)-h_{\text{in}}(v)|<P_{uv},
	\qquad
	\text{harm:}\quad |\gamma_{\text{out}}(u)-\gamma_{\text{in}}(v)|<P_{uv}.
	\]
\end{enumerate}
Successful attempts add a parallel directed edge of the corresponding sign; unsuccessful attempts add nothing. This allows repeated affordances or repeated adverse interactions between the same ordered pair while preventing unbounded multiplicity.

\paragraph{Aging and culling (node removal rule).}
After binding is evaluated for the new node, every node ages by one tick. A node $x$ becomes \emph{eligible} for removal only once its age is at least $L_S$. Eligible nodes are removed (along with all incident edges) if either:
\begin{enumerate}
	\item $x$ has no incoming edges (no support or influence from others), or
	\item the incoming harm fraction exceeds $\eta$:
	\[
	\frac{\mathrm{harm}_{\mathrm{in}}(x)}{\mathrm{help}_{\mathrm{in}}(x)+\mathrm{harm}_{\mathrm{in}}(x)}>\eta.
	\]
\end{enumerate}
Here, \(\mathrm{help}_{\mathrm{in}}(x)\) and \(\mathrm{harm}_{\mathrm{in}}(x)\) count incoming interactions accumulated from subsequently added nodes, retaining those contributions even after the corresponding edges disappear.

Edges do not decay on their own; they disappear only when one endpoint is deleted. Repeating this birth--binding--culling cycle yields the evolving signed network used in subsequent simulations.  We use $\eta=0.5$ unless specified otherwise.

\begin{figure}[H]

	\centering
	\scalebox{0.95}{\begin{tikzpicture}[xscale=10, yscale=10,>=stealth]
			\tikzstyle{v}=[circle, minimum size=1mm,draw,thick]
			\node[v] (A) {$A$};
			\node[v] (B) [right=of A] {$B$};
			\draw [->, blue] (A.20) to   (B.160);
			\draw [->, red ] (B.200) to  (A.340);
	\end{tikzpicture}}\hspace{0.3em}
	\scalebox{0.95}{\begin{tikzpicture}[xscale=10, yscale=10,>=stealth]
			\tikzstyle{v}=[circle, minimum size=1mm,draw,thick]
			\node[v] (A) {$A$};
			\node[v] (B) [right=of A] {$B$};
			\node[v] (C) [above=of B] {$C$};
			\draw [->, blue] (A.20) to   (B.160);
			\draw [->, red] (B.200) to  (A.340);
			\draw [->, red] (C.225) to  (A.45);
			\draw [->, blue] (C.205) to (A.65);
			\draw [->, blue] (C.185) to (A.85);
			\draw [->, red] (B.75) to (C.285);
			\draw [->, blue] (B.105) to (C.255);
	\end{tikzpicture}}\hspace{0.3em}
	\scalebox{0.95}{\begin{tikzpicture}[xscale=10, yscale=10,>=stealth]
			\tikzstyle{v}=[circle, minimum size=1mm,draw,thick]
			\node[v] (A) {$A$};
			\node[v] (B) [right=of A] {$B$};
			\node[v] (C) [above=of B] {$C$};
			\node[v] (D) [right=of B] {$D$};
			\draw [->, blue] (A.20) to   (B.160);
			\draw [->, red] (B.200) to  (A.340);
			\draw [->, red] (C.225) to  (A.45);
			\draw [->, blue] (C.205) to (A.65);
			\draw [->, blue] (C.185) to (A.85);
			\draw [->, red] (B.75) to (C.285);
			\draw [->, blue] (B.105) to (C.255);
			\draw [->, blue] (D.180) to (B.0);
			\draw [->, blue] (D.135) to (C.315);
			\draw [->, blue] (B.340) to (D.200);
	\end{tikzpicture}}

	\vspace{0.5em}
	\scalebox{0.95}{\begin{tikzpicture}[xscale=10, yscale=10,>=stealth]
			\tikzstyle{v}=[circle, minimum size=1mm,draw,thick]
			\node[v] (A) {$A$};
			\node[v] (B) [right=of A] {$B$};
			\node[v, red, dashed] (C) [above=of B] {\textcolor{red}{$C$}};
			\node[v] (D) [right=of B] {$D$};
			\node[v] (E) [above=of D] {$E$};
			\draw [->, blue] (A.20) to   (B.160);
			\draw [->, red] (B.200) to  (A.340);
			\draw [->, red] (C.225) to  (A.45);
			\draw [->, blue] (C.205) to (A.65);
			\draw [->, blue] (C.185) to (A.85);
			\draw [->, red] (B.75) to (C.285);
			\draw [->, blue] (B.105) to (C.255);
			\draw [->, blue] (D.180) to (B.0);
			\draw [->, blue] (D.135) to (C.315);
			\draw [->, blue] (B.340) to (D.200);
			\draw [->, red] (E.270) to (D.90);
			\draw [->, red] (E.160) to (C.20);
			\draw [->, red] (E.200) to (C.340);
	\end{tikzpicture}}\hspace{0.3em}
	\scalebox{0.95}{\begin{tikzpicture}[xscale=10, yscale=10,>=stealth]
			\tikzstyle{v}=[circle, minimum size=1mm,draw,thick]
			\node[v] (A) {$A$};
			\node[v] (B) [right=of A] {$B$};
			\node[v] (D) [right=of B] {$D$};
			\node[v] (E) [above=of D] {$E$};
			\draw [->, blue] (A.20) to   (B.160);
			\draw [->, red] (B.200) to  (A.340);
			\draw [->, blue] (D.180) to (B.0);
			\draw [->, blue] (B.340) to (D.200);
			\draw [->, red] (E.270) to (D.90);
	\end{tikzpicture}}
	
	\caption{Sample simulation showing graph generation and node removal.  Blue arrows are help
		edges and red arrows are harm edges; each arrow points from the acting node to the node
		it affects.  Top row (left to right): Steps 2--4.  Bottom row: Step 5 (E added, C marked
		for removal) and Step 6 (C and its edges removed).  A node is removed when it has no
		incoming edges at all, or when the harm fraction of its incoming edges exceeds $\eta$.
		For clarity the lifespan condition is omitted here; in the full model a node becomes
		eligible for removal only once its age is at least $L_S$.}
	\label{fig:samplesim}
\end{figure}

\subsection{Sample Simulation}
\begin{adjustwidth}{2cm}{2cm}
	\begin{flushleft}
		\textbf{Step 1}: Node A is added.
		
		\textbf{Step 2}: Node B added, harm (red) edge added $B \to A$ as $|\gamma_{\text{in},A}-\gamma_{\text{out},B}| < P_{BA}$, and help (blue) edge added $A \to B$ as $|h_{\text{out},A}-h_{\text{in},B}| < P_{AB}$.
		
		\textbf{Step 3}: Node C added, 3 blue and 2 red links added after threshold test between C and existing nodes A,B.
		
		\textbf{Step 4}: Node D added, 3 blue links added after threshold test between D and existing nodes A,B,C: $D\to B$, $D\to C$, and $B\to D$.

		\textbf{Step 5}: Node E added, 3 red links added. Node C now receives 3 harm edges (from B and E) and only 2 help edges (from B and D), so $\frac{3}{5} = 0.6 > \eta = 0.5$. C is marked for removal (assuming its age exceeds $L_S$). Node D receives 1 help edge (from B) and 1 harm edge (from E), so $\frac{1}{2} = 0.5 \not> \eta = 0.5$; D is not marked for removal.
		
		\textbf{Step 6}: Node C (and all its incident edges) removed. The network continues with nodes A, B, D, E.
	\end{flushleft}
\end{adjustwidth}

\subsection{Eigenvalue Analysis of the Adjacency Matrix}
\label{eigen}
For additional insights into the structure of this adaptive network, we turned to eigenvalue analysis, which provides a higher level description of the simulation from the lower-level data.~\cite{richards2000network}

At each simulation tick, we generate an adjacency matrix by flattening the MultiGraph to a weighted graph by summing the number of edges: +1 for help edges and -1 for harm edges.  Recall that at most 6 edges of one type (help or harm) can form between two nodes. 

For example, in the above Figure \ref{fig:samplesim} sample simulation at \textit{Step 5} (nodes A, B, C, D, E), the signed adjacency matrix $\mathbf{A}_t$ (where $+1$ denotes a help edge and $-1$ a harm edge, summed over parallel edges) is:
\[
\mathbf{A}_t = \bordermatrix{
& A & B & C & D & E\cr
A & 0 & +1 & 0 & 0 & 0\cr
B & -1 & 0 & 0 & +1 & 0 \cr
C & +1 & 0 & 0 & 0 & 0\cr
D & 0 & +1 & +1 & 0 & 0\cr
E & 0 & 0 & -2 & -1 & 0 \cr
}, \quad t=5
\]
Entry $\mat{A}_{ij}$ indicates the net interaction from node $i$ to node $j$. For instance, $\mat{A}_{CA} = +1$ reflects C's net helpful influence on A (2 help edges minus 1 harm edge).

We then approximate the dominant eigenpair (eigenvector and corresponding eigenvalue) $(\hat{\lambda}_1,\vecb{v}_1)$ using a fixed-iteration power method with normalization at each step and a Rayleigh-quotient estimate,

\[
\hat{\lambda}_1 \;=\; \frac{\vecb{v}_1^\top \mat{A}_t \vecb{v}_1}{\vecb{v}_1^\top \vecb{v}_1},
\]

terminating after at most 20 iterations or earlier if successive Rayleigh estimates satisfy $|\hat{\lambda}_{1}^{(k)}-\hat{\lambda}_{1}^{(k-1)}|<10^{-6}$.

To obtain a computationally efficient approximation of the sub-dominant eigenvalue, we perform a rank-one deflation by subtracting the estimated dominant mode,

\[
\mat{B}_t \;=\; \mat{A}_t \;-\; \hat{\lambda}_1\,\vecb{v}_1 \vecb{v}_1^\top,
\]

and reapply the same iterative procedure to $B_t$ to estimate $\hat{\lambda}_2$. We then report the spectral gap as $|\hat{\lambda}_1|-|\hat{\lambda}_2|$ and track the principal-angle change between successive dominant eigenvectors as a proxy for turnover in the network core. Because $A_t$ is generally non-symmetric, these spectral quantities should be interpreted as efficient heuristics for cohesion and regime change rather than as exact spectral characterizations at each time step.

The \textbf{principal eigenvector} for the adjacency matrix encodes the principal or dominant direction of flows or activity in the network.  Its corresponding \textbf{principal eigenvalue} gives a measure to the 'strength' of that dominance. An increasing steady principal eigenvalue also means that average connectivity (and so the potential for feedback/cooperation or harm) is still growing -- the graph hasn't saturated yet.

Our second measure, the \textbf{spectral gap}, is the difference between the two leading eigenvalues: $|\hat{\lambda}_1| - |\hat{\lambda}_2|$. A large and growing spectral gap implies there is one cohesive, resilient strongly connected component.  In other words, the principal direction of connectivity has 'pulled away' from any secondary directions. A small or falling gap would suggest there are many communities and more potential for fragmentation.  Put another way, $\lambda_1$ tells us how \textit{dense} the connections are, the spectral gap tells us how internally cohesive that density is.

Our third measure, the \textbf{principal angle}, reveals how stable the 'identity' of the dominant nodes is.   It measures the cosine similarity between two successive network states:

\[
\theta_t
= \arccos\!\left(
\frac{\vecb{v}_{t-1}^\top \vecb{v}_t}{\|\vecb{v}_{t-1}\|\,\|\vecb{v}_t\|}
\right).
\]

A small, steady angle signals a persistent hub has formed; sharp increases imply regime shifts where new species take over the ‘core’ of the network.

\medskip

\noindent In summary, a high dominant eigenvalue indicates a dense, well-defined pattern of
help and harm interactions; the principal angle measures structural change between successive
network states; and the spectral gap indicates how strongly that primary interaction pattern
dominates.  These measures are collected with the other reported quantities in
Table~\ref{tab:panels}.

What follows are the results from several simulations.  When parameters are altered, they are specifically listed below the figure.

		\section{Simulations \& Results}
		
		In a systematic way, we explore how shifting the parameters (lifespan, 
		harm-to-help ratio, binding chance) influence the emergence and persistence 
		of strongly connected components (SCCs) in our directed network.  Our 
		primary interest was the evolution and robustness of cooperative structures
		(large SCCs) in networks characterized by varying degrees of antagonism
		(harm-to-help ratio).  Eigenvalue analysis was performed to give a higher level
		characterization of the network's behavior in the short- and long-run.

		The central result is that a community of self-interested, randomly
		interacting agents does not need cooperation to be the majority interaction
		type for a large, persistent cooperative core to form. Across
		the parameter sweep, a giant strongly connected component, this model's
		stand-in for a Collective Affordance Set, self-assembles and survives even
		when harmful edges outnumber helpful ones by a wide margin, provided the
		harm-to-help ratio $\rho$ stays below a critical value near
		$\rho_c \approx 0.6$. The component is strongly connected in the full signed
		network, so it is a web of mutual reach rather than a club of mutualists;
		what makes it cooperative is the culling rule, which removes any member past
		its grace period whose incoming harm outweighs its incoming help. Every
		surviving member is therefore helped more than it is harmed, however
		antagonistic the
		interactions on offer. Above the critical value, growth
		halts and the community shifts to a smaller, bounded steady state rather
		than collapsing outright. The binding chance, how readily agents form
		connections at all, changes only \emph{when} this transition happens, not
		\emph{whether} it happens: raising it pulls the transition to smaller
		populations without changing the size the SCC eventually reaches.
		Opportunity to interact governs the timing of collective organization; the
		balance of help to harm governs whether it survives at all.

		This result is deliberately general: nothing in the rules refers to a
		specific organism, metabolite, or ecological guild. But the mechanism it
		describes is not merely an analogy to biology. Autocatalytic sets, the
		motivating example throughout this paper, are themselves a biological
		construct, and the Collective Affordance Set framing makes the same claim
		directly: soil microbial communities assemble stable cooperative cores
		under the same kind of antagonism-dominated conditions studied
		here~\citep{mall2022limited,chang2023emergent,schlatter2019inhibitory}. The
		model is general in scope and biological in application at once.

		The quantities reported in each panel of the simulation figures below are
		defined in Table~\ref{tab:panels}.

		\begin{table}[htbp]
			\centering
			\caption{Quantities reported in the ten panels of
				Figures~\ref{HTH0.3}--\ref{HTH0.7}.  Time-dependent quantities are
				recorded at every step $t$; the degree distributions are taken at the
				end of the run.}
			\label{tab:panels}
			\begin{tabular}{@{}lp{0.60\linewidth}@{}}
				\toprule
				\textbf{Quantity} & \textbf{Definition} \\
				\midrule
				Nodes and edges & Number of nodes and edges present in the graph at
				step $t$.  One node is added per step with its own traits, and edges
				form by the threshold rule.  As in many network growth algorithms,
				incumbents form edges only with the newcomer, yet cycles still
				arise. \\
				\addlinespace
				Fraction of nodes in an SCC & Fraction of nodes at step $t$ belonging
				to any strongly connected component. \\
				\addlinespace
				Fraction of nodes in largest SCC & Fraction of nodes at step $t$
				belonging to the largest strongly connected component. \\
				\addlinespace
				Largest SCC size & Number of nodes in the largest strongly connected
				component at step $t$. \\
				\addlinespace
				In-degree, help and harm & Distribution of the number of inbound help
				and harm edges at the end of the run. \\
				\addlinespace
				Out-degree, help and harm & Distribution of the number of outbound
				help and harm edges at the end of the run. \\
				\addlinespace
				Mean dominant eigenvalue & Density of connections in the direction of
				the principal eigenvector. \\
				\addlinespace
				Absolute mean dominant eigenvalue & The same quantity with the sign
				removed, isolating the intensity of shifts in the principal
				eigenvector. \\
				\addlinespace
				Principal angle & Angle between the principal eigenvectors of
				successive network states. \\
				\addlinespace
				Spectral gap & Difference between the first and second principal
				eigenvalues, $|\hat{\lambda}_1| - |\hat{\lambda}_2|$. \\
				\bottomrule
			\end{tabular}
		\end{table}

		\subsection{Harm-To-Help Ratio and Lifespan Dynamics}
		
		\subsubsection{Low antagonism ($\rho$ below $~0.5$) }
		When cooperative interactions dominate (harm ratio $\rho< 0.5$), we observed 
		exponential growth in directed edge formation, resulting in a linearly 
		increasing 
		SCC size relative to the number of nodes in the network.  This scenario 
		characterizes a cooperative and expansive community where nodes not only 
		persist, 
		but integrate rapidly into a collective structure. Once such a community 
		emerges, 
		with a majority of help edges, it would be difficult for its elements to be 
		removed. 
		
\begin{figure}
			\centering
				\includegraphics[width=\linewidth,height=0.76\textheight,keepaspectratio]{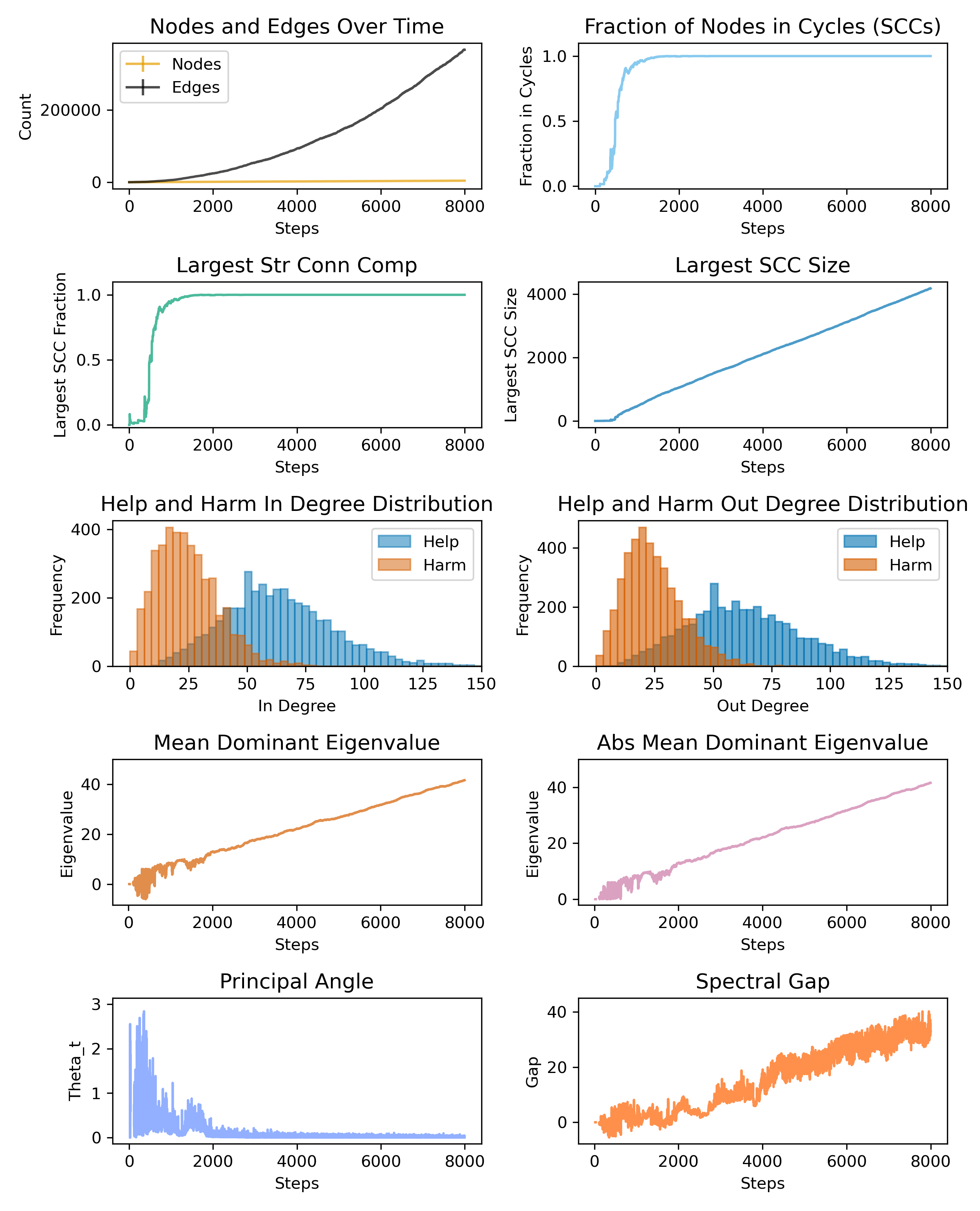}
				\caption{Simulation summary at harm-to-help ratio $\rho = 0.3$, node
					lifespan $L_S = 100$, and binding chance $p_{max} = 0.003$, over
					8,000 steps.  Panels, by row: node and edge counts over time; the
					fraction of nodes belonging to any strongly connected component
					(SCC); the fraction of nodes in the largest SCC; the size of the
					largest SCC; in-degree and out-degree distributions of help (blue)
					and harm (red) edges at the end of the run; the mean dominant
					eigenvalue of the adjacency matrix and its absolute value; the
					principal angle between successive network states; and the spectral
					gap.  With cooperation dominant, edges accumulate superlinearly,
					essentially every node joins a single SCC within about 1,500 steps,
					help degrees sit well above harm degrees, and the principal angle
					decays toward zero as a stable dominant interaction pattern
					establishes itself.}
				\label{HTH0.3}
			
		\end{figure}

\begin{figure*}
			\centering
				
				\includegraphics[width=\linewidth,height=0.76\textheight,keepaspectratio]{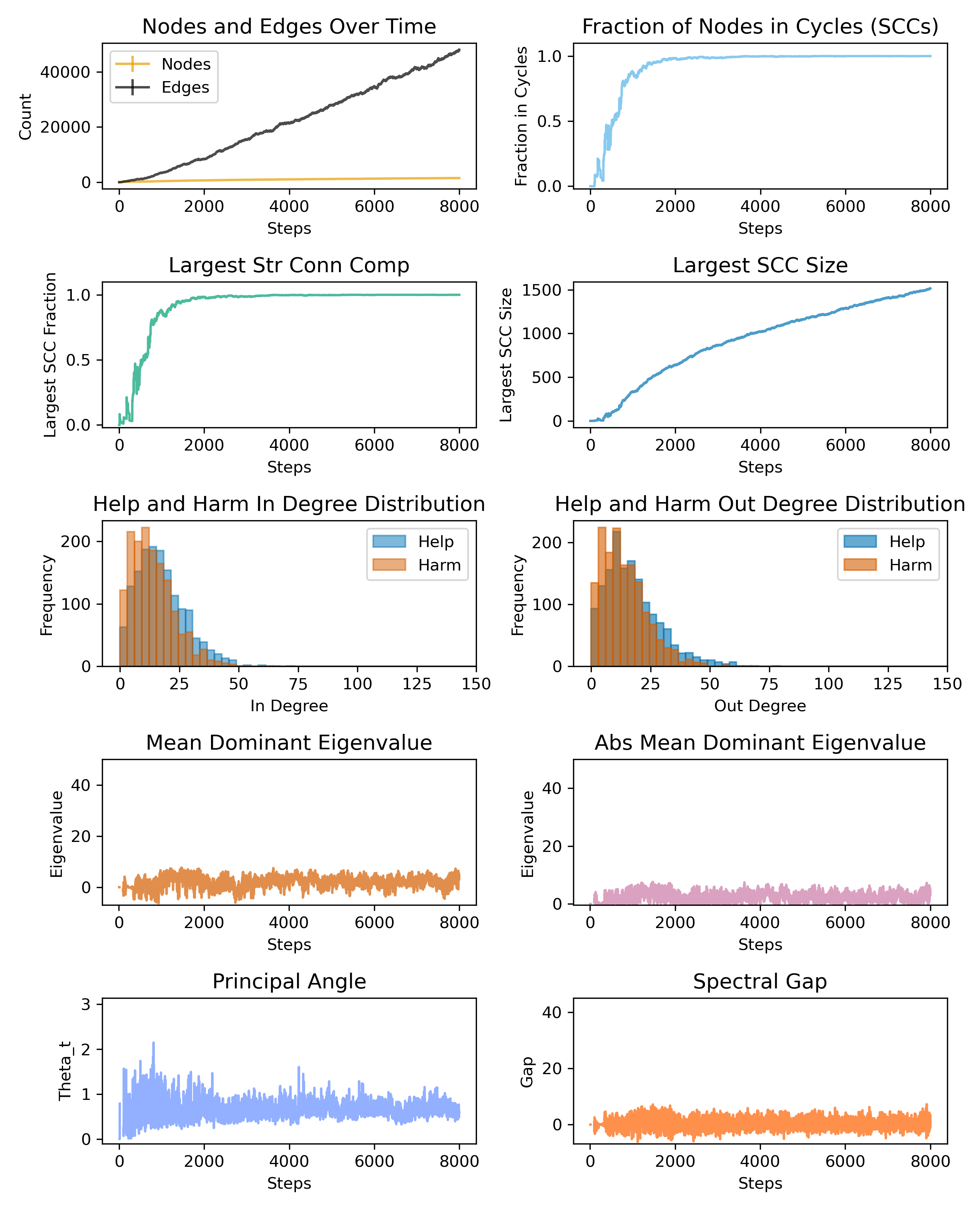}
				\caption{Simulation summary at harm-to-help ratio $\rho = 0.5$, with
					$L_S = 100$, $p_{max} = 0.003$, and the same ten panels as
					Figure~\ref{HTH0.3} (node and edge counts, SCC membership and size,
					help and harm degree distributions, the two dominant-eigenvalue
					measures, principal angle, and spectral gap).  At parity between
					help and harm, edge growth is linear rather than superlinear.  The
					largest SCC still absorbs essentially the entire population, but
					accumulates more slowly and reaches roughly 1,500 nodes by step
					8,000.  Help and harm degree distributions now overlap almost
					completely at low degree, the dominant eigenvalue and spectral gap
					remain near zero, and the principal angle continues to fluctuate
					rather than settling: the network is fully connected yet
					structurally unsettled.}
				\label{HTH0.5}
		\end{figure*}

\begin{figure*}
			\centering

				\includegraphics[width=\linewidth,height=0.76\textheight,keepaspectratio]{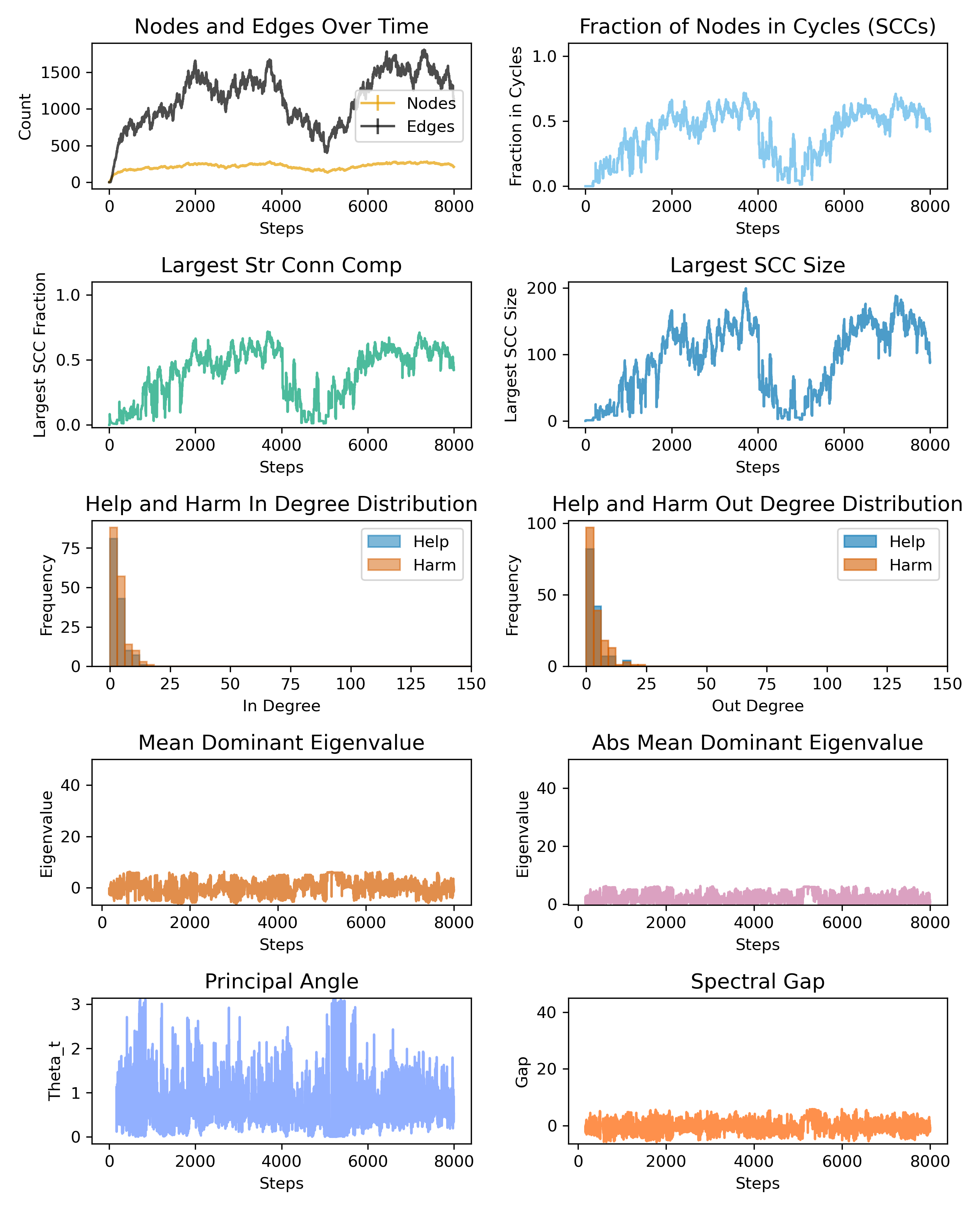}
				\caption{Simulation summary at harm-to-help ratio $\rho = 0.7$, with
					$L_S = 100$, $p_{max} = 0.003$, and the same ten panels as
					Figure~\ref{HTH0.3} (node and edge counts, SCC membership and size,
					help and harm degree distributions, the two dominant-eigenvalue
					measures, principal angle, and spectral gap).  Under strong
					antagonism the population no longer grows: node and edge counts
					fluctuate about a bounded level, and the largest SCC oscillates
					between roughly a tenth and two thirds of the population, including
					a pronounced collapse and recovery near step 5,000.  Help and harm
					degree distributions are nearly indistinguishable and confined to
					low degree, the dominant eigenvalue and spectral gap sit near zero,
					and the principal angle stays large and erratic, indicating that no
					stable dominant interaction pattern is ever established.  A
					strongly connected core nevertheless persists, continually rebuilt
					from newly added nodes.}
			\label{HTH0.7}
		\end{figure*}

		Figures \ref{HTH0.3},\ref{HTH0.5},\ref{HTH0.7} show the results of the 
		simulation from solely increasing the Harm-To-Help ratio from $\rho =0.3, 
		0.5, 0.7$ with all other parameters fixed.  At first, we can restrict our 
		view to the differences in the number of nodes and edges at step $t$, the 
		largest SCC size and largest SCC fraction.

		\subsubsection{High antagonism ($\rho$ above $~0.5$)}
		
\begin{figure}[H]
			\includegraphics[width=\linewidth]{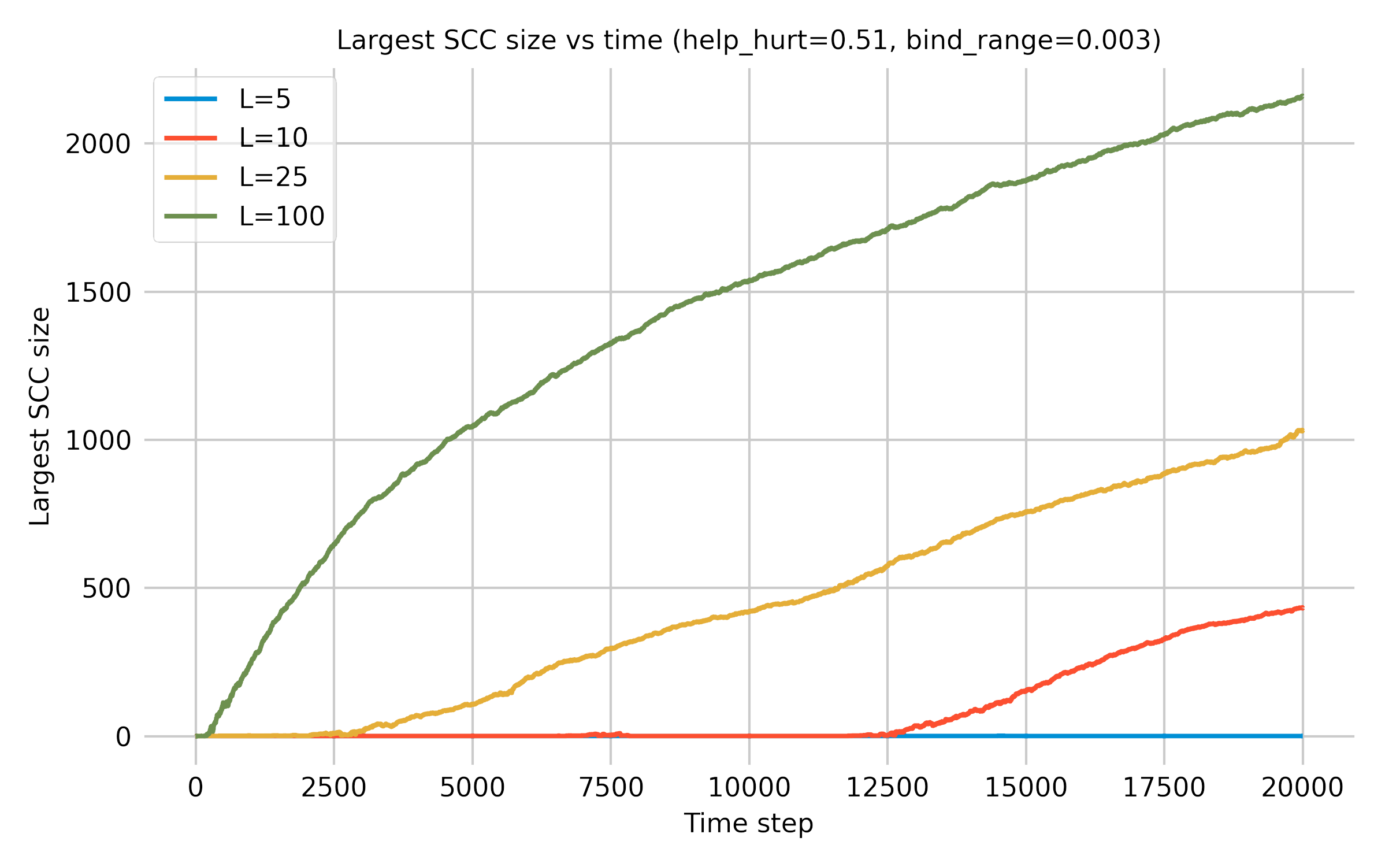}
			\caption{Growth of the largest strongly connected component (SCC) over
			time at a harm-to-help ratio just above parity ($\rho = 0.51$) and binding
			chance $p_{max}=0.003$, for four node lifespans $L_S$.  At $L_S=5$ no SCC
			ever forms.  At $L_S=10$ one appears only after roughly 12,500 steps and
			remains small, whereas $L_S=25$ and $L_S=100$ produce components that
			emerge much earlier and continue growing throughout the run.  Once
			antagonism exceeds parity, node lifespan governs whether a persistent
			cooperative core can form at all.}
			\label{lifespan_rho}
		\end{figure}

		For harm-to-help ratios exceeding 0.5, a strongly connected community still 
		emerges 
		and persists, provided nodes' lifespan surpasses $\approx 10$ timesteps.  
		Below this 
		number, an SCC will fail to form reliably if $\rho$ is sufficiently high.  
		See 
		Figure \ref{lifespan_rho}. 
		
		Notably here, SCC size does not grow indefinitely; rather, it reaches an 
		upper bound and fluctuates around a steady-state maximum. This equilibrium 
		state depends directly on both lifespan and the binding chance parameters. 
		Increasing antagonism (harm ratio) gradually reduces the steady-state size 
		of the SCC, yet even with extremely antagonistic environments (harm ratio 
		approaching 0.9), some significant level of persistent inter-connectedness 
		remains, as tenuous as it is.


		\subsection{Influence of Binding Chance}
		
		Adjusting the binding chance, $p_{max}$ which scales the likelihood 
		threshold for 
		interactions, primarily impacts the rate of convergence toward the final 
		dynamic 
		state (Figure \ref{transition}).  Increasing this parameter leads more 
		quickly to 
		equilibrium behaviors without substantially altering the qualitative 
		behavior of SCC 
		growth.  
		
		However, in scenarios where the harm ratio exceeded 0.5, a higher binding 
		chance 
		increased the \textit{fraction} of nodes in the SCC, but overall slightly 
		decreased 
		the number of nodes in the equilibrium size of the SCC.  This suggests that 
		increasing the likelihood of interaction partially counteracts the 
		destabilizing 
		effect of widespread antagonism.  See Table~\ref{pmaxtable}.

		\begin{table}[htbp]
			\centering
			\caption{Effect of the binding chance $p_{max}$ on the largest strongly
				connected component under strong antagonism, measured after 30,000
				steps with node lifespan $L_S = 100$ and harm-to-help ratio
				$\rho = 0.7$.  Raising $p_{max}$ draws a progressively larger
				\emph{fraction} of the surviving population into the component while
				its absolute \emph{size} falls, so more frequent interaction
				opportunity buys near-complete connectivity within a smaller standing
				population.}
			\label{pmaxtable}
			\begin{tabular}{c c c}
				\toprule
				$p_{max}$ & Fraction in SCC & Final SCC Size \\
				\midrule
				0.005 & $\approx$0.8 & $\approx$210 \\
				0.0075 & $\approx$0.9 & $\approx$200 \\
				0.015 & $\approx$0.98 & $\approx$175 \\
				0.02 & $\approx$1.0 & $\approx$150\\
				\bottomrule
			\end{tabular}
		\end{table}

		\subsection{Node Age and Persistence}
		
		We noted a `demographic' shift among highly-connected nodes (in the top 10\% 
		of in-degree) as antagonism increased.  Specifically, and perhaps not 
		surprisingly, networks with higher antagonism were increasingly dominated by 
		younger nodes.  High-harm environments shorten the node survival 
		significantly, reducing the presence of ``older" and more established 
		nodes.  Despite this shorter persistence, large SCCs remain viable and 
		indicate a dynamically renewing structure of cooperation, which is resilient 
		amidst high node turnover.

\begin{figure}
			\begin{center}
				
				\includegraphics[width=0.9\linewidth]{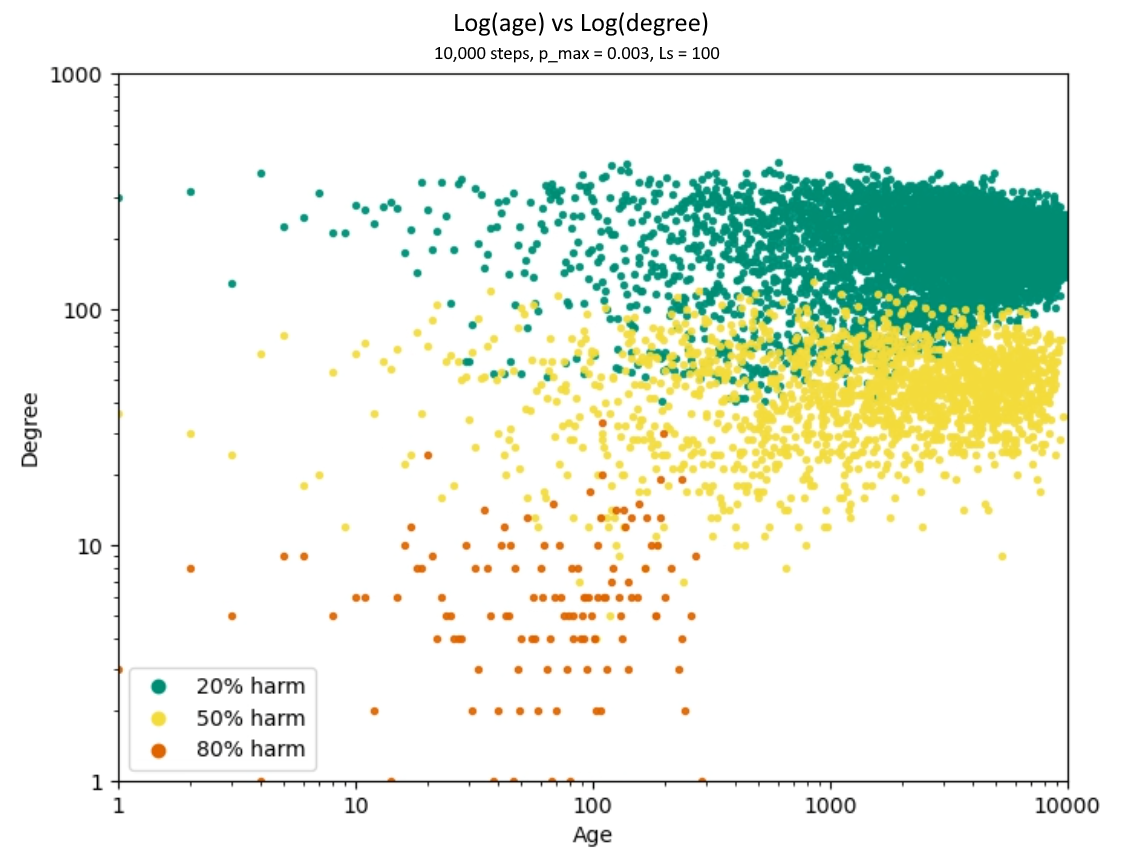}
				\caption{Node age versus total degree at the end of a 10,000-step
					simulation, both axes logarithmic.  Each point is one node in a
					network of $N=10{,}000$ with binding chance $p_{max}=0.003$, shown
					at three levels of antagonism: $\rho=0.2$, $0.5$, and $0.8$.  At low
					antagonism, old and highly connected nodes form visible bands,
					reflecting an established core.  These bands dissipate as $\rho$
					increases and the population shifts toward younger nodes; at
					$\rho=0.5$ a distinct cluster persists for most of the simulation
					but at reduced degree.  A time-lapse animation of this distribution
					developing over the course of a simulation is provided as
					Supplementary Video~S1.}
				\label{scatter1}
			\end{center}
		\end{figure}

		When viewing log(age) vs log(degree) we see bands or clustering when 
		Harm-To-Help is $< 0.5$ which dissipates and trends towards younger nodes as 
		the network is stressed by harmful interactions.  See Figure \ref{scatter1}.
		
		\section{Discussion}
		
		\paragraph{Clarifying Scope of Vanilla Model}
		The framework introduced here operates deliberately at a higher 
		level of 
			abstraction. Rather than modeling specific biochemical, ecological, or 
			evolutionary 
			mechanisms that generate interactions (e.g., metabolic exchange, spatial 
			structuring, gene loss, or adaptive specialization), the model treats 
			the 
			interaction network itself as the primary object of study. In this 
			sense, it 
			explores the space of possible help--harm configurations that such 
			mechanisms may 
			produce, rather than the mechanisms themselves. This abstraction allows 
			us to 
			identify network-level regularities—such as the persistence of large 
			strongly 
			connected components under mixed antagonism and support—that are 
			potentially shared 
			across microbial, economic, and other complex systems. At the same time, 
			this 
			perspective necessarily leaves unresolved questions about how particular 
			interaction 
			patterns arise, how they are modulated by spatial or metabolic 
			constraints, or how 
			they evolve over long timescales. The present results should therefore 
			be 
			interpreted as constraints on, rather than substitutes for, more 
			detailed 
			mechanistic models: any system whose interaction structure falls within 
			this class 
			would be expected to exhibit similar collective behavior, regardless of 
			the 
			underlying generative process.
		
		\subsection{Robustness of SCCs under High Antagonism}
		
		Our model demonstrates robustness of large SCC structure
		even when 
		antagonistic interactions (harmful edges) significantly outnumber supportive 
		interactions. This finding could be an extension of the concept of `islands 
		of 
		cooperation' in \citep{nowak1999games}.  Specifically, even with high 
		harm-to-help 
		ratios ($\rho > 0.5$), SCCs consistently emerge and persist at steady-state 
		equilibria. This implies the presence of inherent resilience within the 
		network 
		structure, which allows it to dynamically reconfigure to maintain 
		cooperative 
		viability despite continuous antagonistic pressure.  The results on node 
		``age" vs 
		degree distribution lend another perspective on the long run dynamics, which 
		we 
		discuss later.
		
		The interaction structure considered here assumes that benefits 
		and harms 
			are effectively public, an assumption that simplifies analysis but
			departs from many
			microbial systems in which goods are partially privatized or retained to
			varying
			degrees. In natural communities, metabolites may be incompletely shared,
			spatially
			constrained, or dynamically regulated, producing interaction strengths
			that vary
			over time and context.

			A related simplification is that our model resolves interactions
			pairwise, while community-level outcomes are frequently not reducible to
			pairwise effects. How far coexistence in diverse communities turns on effects
			that are not reducible to pairs is a developed body of theory, covering both
			the mechanisms by which higher-order effects alter
			coexistence~\citep{levine2017beyond} and the conditions under which they change
			the outcome in a diverse community~\citep{gibbs2022coexistence}. In a
			seven-member synthetic bacterial community on
			duckweed, whose pairwise interactions were mostly negative, pairwise
			cocultures did not predict the assembled community; adding three-member
			combinations recovered both the observed structure and the interaction
			coefficients, with the higher-order terms damping the strongest pairwise
			effects and so preventing overabundance and
			extinction~\citep{ishizawa2024beyond}. Our model has no such terms, and
			adding them is a natural extension. It does, though, supply a null case
			worth having when results of that kind are interpreted: here a
			community-level cooperative core coexists with predominantly antagonistic
			pairwise ties in a network built from pairwise rules alone, because
			turnover removes the members that antagonism dominates. A mismatch between
			a community's pairwise interaction census and its realized structure is
			thus not on its own evidence of higher-order interactions.
		
		The stability of these SCCs under high antagonism can be explained in 
		network-theoretic terms: while harmful edges break node support structures, 
		nodes 
		with multiple positive inbound edges form resilient core communities, even 
		when 
		positive edges have a low probability of being formed. This leads to a 
		dynamic 
		steady state characterized by continuous recruitment and turnover of 
		peripheral 
		nodes, rather than collapse of the entire structure. These results align 
		conceptually with robustness observed in dynamic real-world networks under 
		external 
		stresses, pointing to perhaps a broader principle of cooperative resilience 
		inherent 
		to certain threshold-based network structures.

\subsection{Thresholds and Regime Shifts}
		
\begin{figure}[H]
			\centering
			
			\includegraphics[width = 0.7\linewidth]{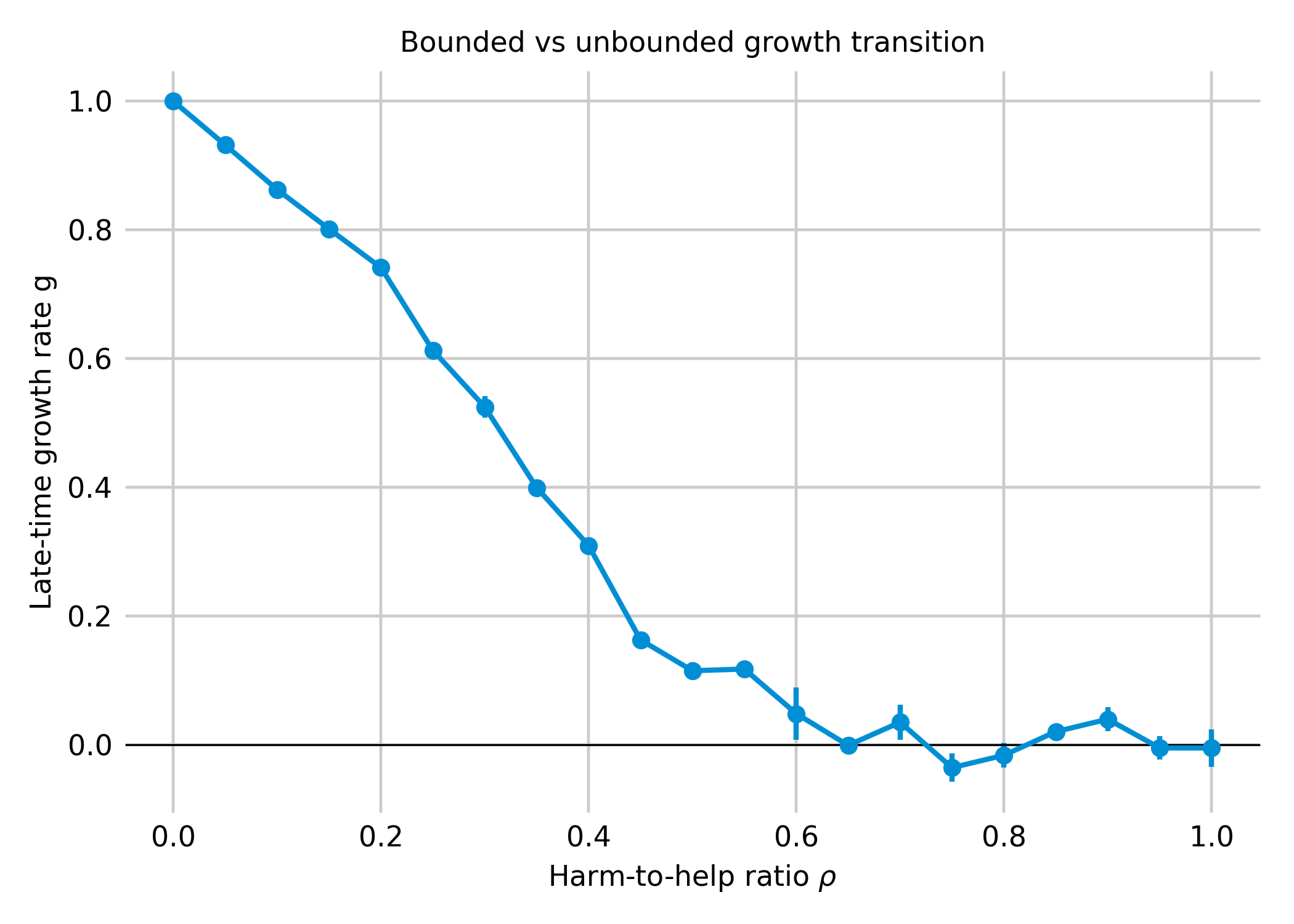}
			\caption{The late-time (last 500 steps) population growth 
			rate $g$ as 
					a function of the harm-to-help ratio $\rho$.  For small $\rho$, 
					the system 
					exhibits sustained positive growth. As $\rho$ increases, the 
					growth rate 
					decreases and crosses zero near $\rho_c \approx 0.6$, 
					marking a shift between distinct dynamical regimes, from sustained 
					growth to a bounded population regime.}

			\label{fig:lategrowthvsrho4}
			
		\end{figure}

\begin{figure}[H]
			\centering
			
			\includegraphics[width = 0.7\linewidth]{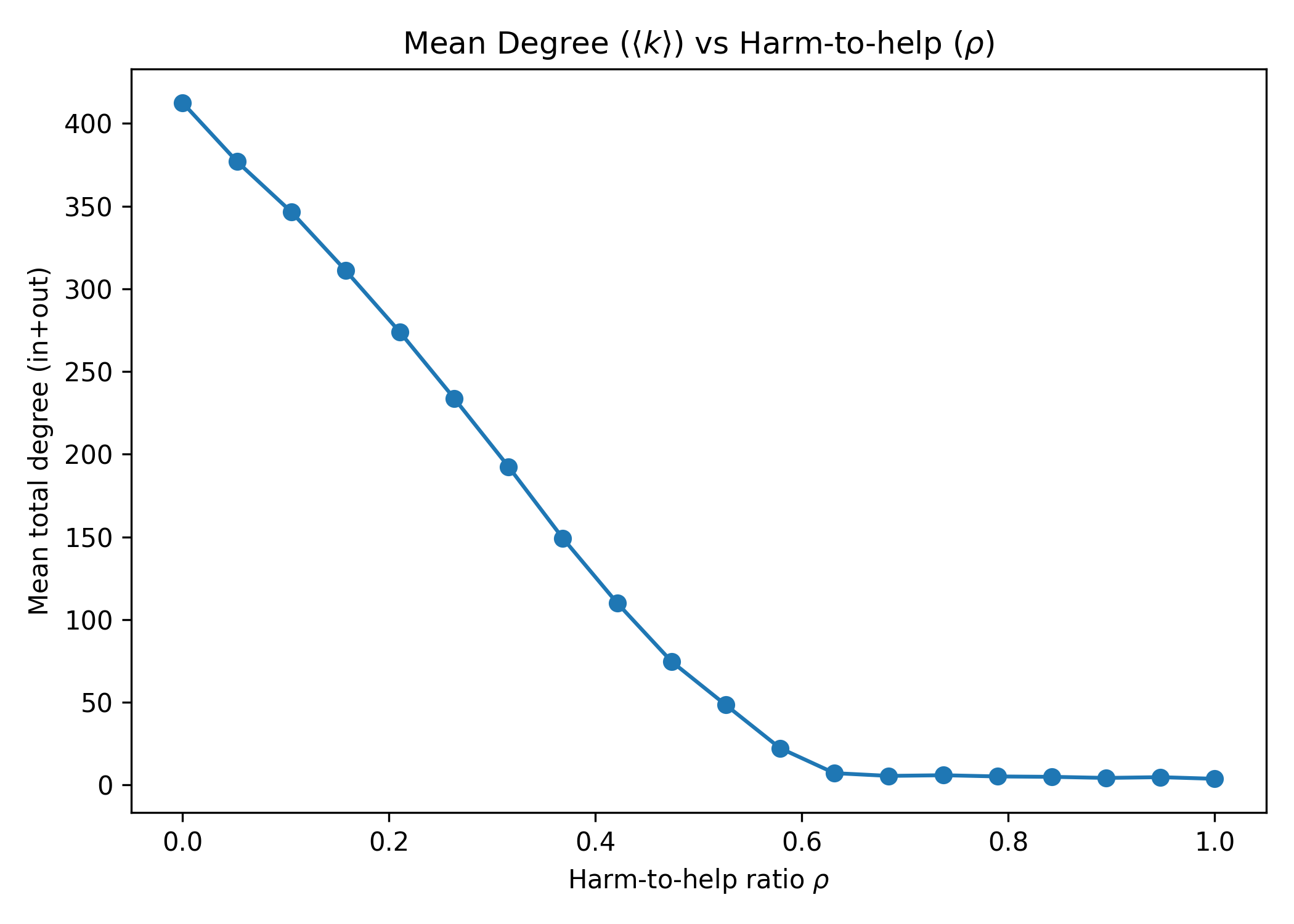}
			\caption{Mean total degree (in + out) as a function of the 
					harm-to-help ratio $\rho$.  As $\rho$ increases, the average 
					connectivity of the 
					network decreases steadily, reaching its floor 
					near $\rho_c 
					\approx 0.6$.  This loss of interaction density coincides 
					with the 
					crossover from sustained to bounded population growth observed 
					in Figure \ref{fig:lategrowthvsrho4}, suggesting the loss of 
					sufficient interaction capacity 
					underlies the 
					shift between dynamical regimes in population behavior.}
			\label{fig:mean_rho}
		\end{figure}

			A central outcome of our simulations is the emergence of a regime change near the harm-to-help ratio $\rho_{c} 
			\approx 
			0.6$. (Figure 
		\ref{fig:lategrowthvsrho4}, \ref{fig:mean_rho}) For values of 
		$\rho$ 
			below this 
			threshold, the strongly connected component (SCC) grows steadily with 
			the total node 
			population, reflecting a regime of sustained interaction in which help 
			predominates 
			among the survivors. Above this threshold, however, SCC growth transitions from 
			indefinite 
			expansion to a bounded, fluctuating steady state. 
			
			This transition is evident in the late-time population growth rate, which decreases as $\rho$ increases and crosses zero near $\rho_c \approx 0.6$ (Figure \ref{fig:lategrowthvsrho4}). We define the late-time growth rate $g$ as the average per-step change in population size over the final $W=500$ time steps:
			\[
			g \;=\; \frac{|V_T| - |V_{T-W}|}{W}.
			\]
			Mean network degree declines over the same range and reaches its floor near the same value (Figure \ref{fig:mean_rho}), a loss of interaction density that accompanies the change in growth dynamics.  Neither curve is discontinuous at $\rho_c$: both fall smoothly and almost linearly across the sweep.  What changes at $\rho_c$ is not the slope but the sign of $g$, and with it the regime.  Below $\rho_c$ the population and its cooperative core expand without bound; above it they do not.

\begin{figure*}
			\centering
			\includegraphics[width=0.9\linewidth]{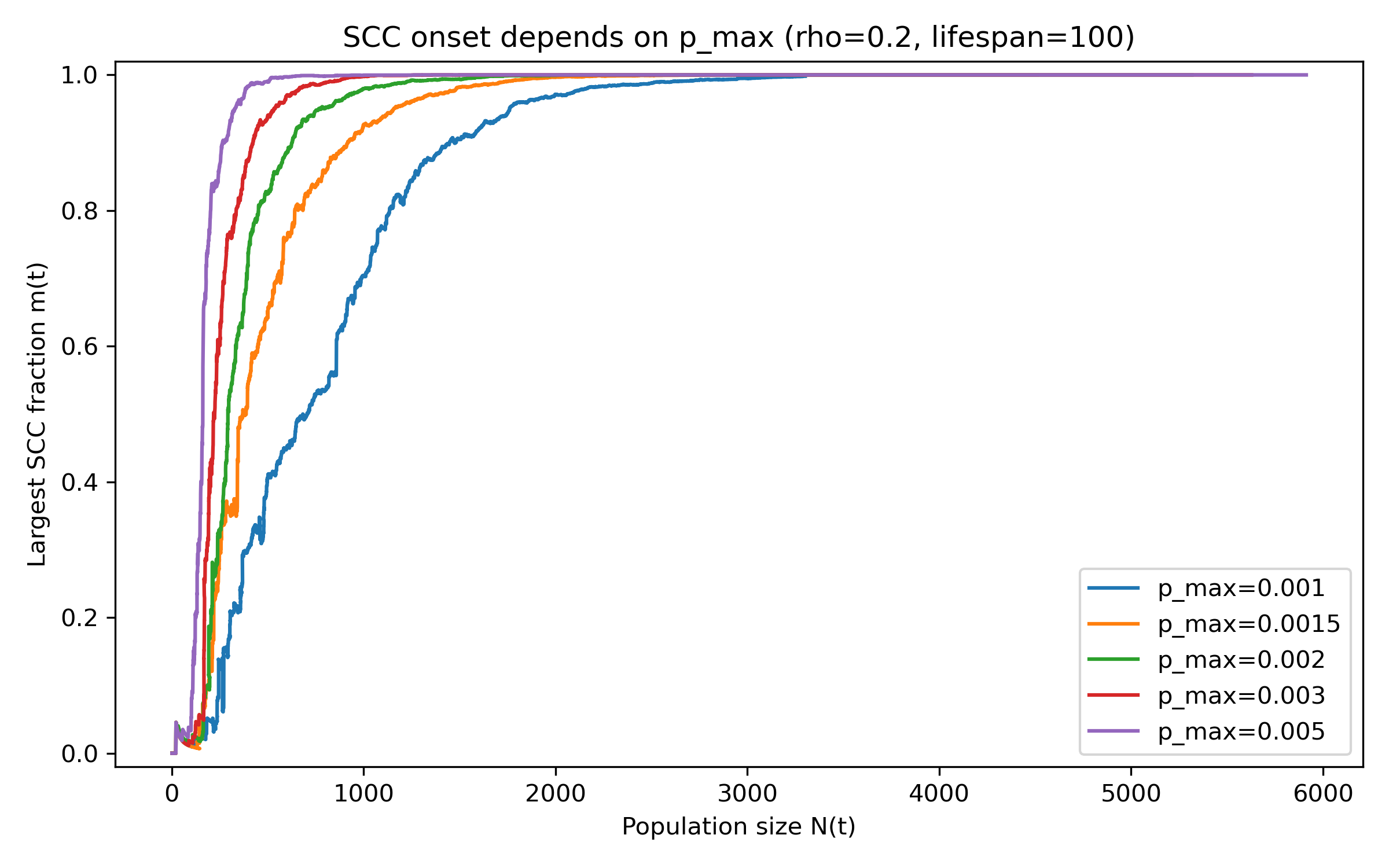}
			
			\caption{Onset of the largest strongly connected component 
			(SCC) as a 
					function of population size for varying bind chance $p_{max}$ 
					($\rho = 0.2$, 
					$L_S=100$). Increasing $p_{max}$ shifts the emergence of global 
					connectivity to 
					progressively smaller population sizes, indicating a transition 
					in the onset of 
					coordination.  All curves converge to a near-unity SCC fraction, 
					showing that 
					$p_{max}$ primarily controls the timing of collective 
					organization.}
			\label{transition}
		\end{figure*}

		While the harm-to-help ratio $\rho$ determines whether the system 
			exhibits sustained or bounded growth, the binding chance $p_{max}$ 
			governs the scale 
			at which global coordination emerges. Figure \ref{transition} 
			shows the 
			fraction of nodes in the largest strongly connected component as a 
			function of 
			population size for different values of $p_{max}$.  Increasing $p_{max}$ 
			reduces the population scale required for SCC formation, 
			indicating a 
			change in coordination regime onset rather than final network structure.  
			
			This behavior is reminiscent of the connectivity thresholds in percolation and 
			random graph models, in which a
			``giant'' component becomes likely once the effective connection density is sufficiently high.~\citep{erdos_renyi_1960}
			In our finite, stochastic setting the onset is rounded, but increasing 
			$p_{max}$ consistently shifts
			coordination to smaller population sizes. Together, these results indicate that $\rho$ governs the
			growth--boundedness regime, whereas $p_{max}$ governs the population 
			scale at which coordinated SCC structure typically emerges.

		These thresholds offer concrete and testable predictions about the 
		conditions required for cooperative network formation and may be connected 
		to critical phenomena observed in physical and complex systems.

		\subsection{Eigenvalue Analysis}
		
		Eigenvalue analysis of the network’s adjacency matrix provided us more insight 
		into 
		structural and dynamical features. The principal eigenvalue (approximated by 
		method 
		outlined in Section \ref{eigen}) measures overall connectivity strength 
		along the 
		dominant eigenvector direction, which reflects the magnitude and stability 
		of 
		cooperation (or antagonism) in the network. Higher eigenvalues indicate 
		denser and 
		strongly coupled cooperative or competitive interactions. In our 
		simulations, rising 
		mean eigenvalues (observed with $\rho < 0.5$ ) corresponded to increasingly 
		stable 
		and interconnected SCCs.  
		
		The spectral gap (difference between the two largest eigenvalues) serves as 
		a helpful secondary indicator, measuring how distinctly the principal `mode' 
		of interaction separates from other competing modes. A widening gap implies 
		a clearly dominant, cohesive community structure, whereas narrowing gaps 
		suggest fragmentation, multiple competing SCCs, or potential instability. 
		Our results consistently showed that networks with high cooperation exhibit 
		both large eigenvalues and pronounced spectral gaps, reaffirming the 
		stability and dominance of a singular cooperative community structure.
		
		Finally, the principal angle between consecutive eigenvectors quantifies the 
		stability of core network structure across simulation steps. Small and 
		stable principal angles indicate persistent dominance by specific nodes or 
		clusters, while sudden spikes highlight structural rearrangements or shifts 
		in network “leadership.” This measure gives additional, nuanced insights 
		into temporal stability and dynamic structural shifts within the network.  
		Consistent with our other measures, when the harm ratio was high, the 
		principal angle fluctuated with increasing magnitude and signaled 
		substantial shifts of influence of the key nodes in the support structure.

		\subsection{Age Distribution and Community Stability}
		
		The interplay between node lifespan and network antagonism produces 
		distinctive 
		patterns in node age (measured in time - birth-step) distributions (Figure 
		\ref{scatter1}). Under conditions of low antagonism (low $\rho$), older 
		nodes with 
		high connectivity become established, reflecting stable core communities. 
		Conversely, increased antagonism rapidly shifts this balance toward younger 
		nodes, 
		as frequent node removal due to harmful interactions reduces node longevity. 
		Interestingly, despite the reduction in older, established nodes, SCCs 
		remain viable 
		by continuously replenishing their core through newly added nodes.  This 
		activity 
		reflects a structurally dynamic equilibrium rather than a fixed, static 
		composition.  Investigating the details behind the bands and clusters we see 
		in the 
		logarithmic plot may yield more insights.  It is notable that at $\rho=0.5$ 
		we see a 
		cluster of nodes that are persisting most of the simulation, but have a 
		reduced 
		overall degree.
		
		This age-based dynamic provides additional perspective on network stability 
		and resilience. The continual integration of younger nodes suggests that 
		organization under antagonism relies not on individual node longevity, but 
		the network’s capacity for rapid turnover and ongoing renewal. This could 
		mirror adaptive responses observed in other complex systems where dynamic 
		reconfiguration rather than static preservation promotes robustness.
		
		A time-lapse animation of this turnover, showing the joint distribution of
		node age and degree developing over the course of a simulation, is provided
		as Supplementary Video~S1.
		
		\subsection{Binding Range as Ecological Analog}
		The binding range parameter, which can be thought of as the probability or 
		``ease" of 
		forming interactions (akin to biochemical or ecological compatibility), 
		accelerates 
		the dynamics of community structuring.  It is interesting this parameter 
		does not 
		radically alter long-term outcomes but modulates the speed and slightly 
		reduces the 
		SCC's ceiling under antagonistic conditions.  In ecological contexts, this 
		could 
		suggest ecosystems with higher interaction potentials (dense signaling, 
		physical 
		proximity, etc) might achieve stable cooperation more rapidly and sustain 
		slightly 
		larger cohesive structures despite competitive pressures. Nevertheless, even 
		in 
		cases with a low binding range, it is ``just a matter of time'' before 
		cooperative 
		communities (autocatalytic sets~\citep{farmer1986autocatalytic}, 
		ecosystems (biological, economic, cultural), etc.) 
		will be formed.

		\subsection{Emergent Levels of Organization in Diverse Systems}\label{sec:diverse}
		
		Our adaptive threshold network model, while originally inspired by microbial 
		community dynamics, embodies general principles of emergent cooperation and 
		the formation of higher-level organizational structures. The minimal 
		assumptions we built into the model make it applicable to a wide variety of 
		complex systems.
		
		The dynamics observed here resonate strongly with other computational works 
		~\citep{alakuijala2024computational} and the concept of collective 
		affordance 
		sets~\citep{kauffman2023third}: where diverse system elements spontaneously 
		combine their causal properties to create new functional wholes. Our 
		findings illustrate clearly that, given sufficient time and diversity of 
		interactions, and provided antagonism stays below the critical ratio, these 
		self-sustaining organizational structures are not merely possible but can be 
		robustly expected to emerge. 
		
		Similarly, ecosystems more broadly demonstrate these principles. Rich 
		biodiversity provides a palette of interactions, allowing ecosystems to 
		self-organize into resilient structures such as trophic networks, 
		mutualistic partnerships, and adaptive community assemblies. Our work aligns 
		with classical general systems theory~\citep{von1968general} and recent 
		empirical studies demonstrating self-organizing structures in ecological 
		contexts, such as microbial communities~\citep{faust_raes_2012} and 
		biofilms~\citep{flemming2016biofilms}.
		
		Biofilms represent another powerful biological parallel. Individual 
		bacterial species, each with unique traits and potentially antagonistic 
		interactions, reliably assemble into complex multicellular aggregates. The 
		emergent properties of biofilms such as shared resource utilization, 
		collective defense mechanisms, and coordinated behavior, demonstrate exactly 
		the kind of spontaneous higher-level organization our model predicts. 
		Indeed, such structures can be viewed as an evolutionary step toward 
		multicellularity~\citep{flemming2016biofilms,claessen2014bacterial,Bozdag2023}.
		
		Placing our model within the broader framework of major evolutionary 
		transitions~\citep{smith_szathmary_1995,lehman_kauffman_2021} also yields 
		valuable insights. It highlights conditions under which Darwinian 
		preadaptations can coalesce into higher-order systems, suppressing or 
		transcending individual-level competition. So, our results offer a 
		theoretical foundation for understanding transitions from simple chemical 
		reaction networks to living cells, from unicellular organisms to 
		multicellular structures, from individuals to 
		groups~\citep{Garcia-Rodriguez2024}, and from isolated economic agents to 
		integrated economies.
		
		In economic networks, analogous dynamics are not hard to find. Various 
		goods, services, technologies, and businesses interact through 
		interdependent relationships which support or undermine one another and give 
		rise to intricate webs of economic cooperation. These adaptive networks 
		continuously reorganize as new elements appear and obsolete ones exit, 
		mirroring our model’s node addition and culling dynamics. Through this lens, 
		our model can offer quantitative insights into the conditions under which 
		economic ecosystems reliably generate new collective structures, whether 
		clusters of industry specialization, innovation hubs, or resilient supply 
		chains. 
		
		By demonstrating robust self-organization through minimal assumptions, our 
		adaptive threshold network model not only sheds light on microbial dynamics 
		but provides a versatile conceptual tool applicable across biological, 
		economic, and social domains. Given enough time, diversity, and interaction 
		potential, the emergence of cooperative structures and new levels of 
		organization is not merely possible but expected, within the bounds the model 
		identifies.

		\section{Conclusions and Future Work}

		Under minimal assumptions, adaptive threshold networks give rise to 
		cooperative communities that persist in the presence of substantial 
		antagonism. The condition is quantitative and specific: a large, strongly 
		connected core assembles and survives as long as the harm-to-help ratio stays 
		below a critical value near $\rho_c \approx 0.6$, and above that value the 
		community does not collapse but stops growing, holding a bounded size while 
		its membership turns over.

		For microbial ecology this reframes a familiar puzzle. A community whose 
		pairwise interactions are mostly antagonistic, which is what several soil 
		surveys report~\citep{foster2012competition,wang2022substrate,schlatter2019inhibitory}, 
		is not for that reason a community that should fall apart. What its members 
		need is not that help be common, but that each of them receive more help than 
		harm, and the turnover that removes those who do not is the same process that 
		keeps the core intact. The two parameter-dependent regime shifts identified in 
		the model suggest that at some point quantity turns into quality. Our results 
		provide support for the conjecture that life inherently "finds a way," 
		tinkering to build cooperative structures despite adverse conditions.
		
		Our current model intentionally excludes stronger forms of evolutionary and 
		co-evolutionary dynamics, setting these aspects as key areas for future 
		development. Integrating evolutionary games --- which have been widely 
		explored 
		elsewhere~\citep{smith_szathmary_1995,nowak_2006_five,Santos2008} --- 
		could enrich our understanding of how cooperation not only emerges but is 
		subsequently refined and stabilized through selective pressures.
		
		Another aspect of our model that remains to be explored is the fact that it 
		can be seen as an example of strong 
		emergence~\citep{BarYam:2004,bedau2008emergence,Schmickl2022,10.1162/artl_a_00397}
		 and downward 
		causation~\citep{Campbell1974,Bitbol2012,farnsworth_ellis_jaeger_2017,Flack2017}.
		 Once communities are formed (at a higher scale), they influence the 
		survivability of nodes (at a lower scale). Thus, it is not possible to 
		reduce the behavior of the system to the properties of its elements. 
		Nevertheless, there are epistemological nuances that should be considered 
		and are beyond the scope of this paper.
		
		Furthermore, the notion of resources is presently implicit in the parameter 
		`lifespan,' which dictates how long a node persists without adequate 
		support. Introducing explicit resource dynamics and spatial constraints 
		would enhance biological realism, enabling the exploration of more nuanced 
		ecological and economic scenarios. Spatial structure, in particular, can 
		affect opportunities for interaction and substantially impact emergent 
		dynamics.~\citep{hauert2004spatial,perc2013collective}

		A natural step is to identify biologically and economically realistic 
		parameter ranges, particularly for lifespan and interaction (binding) 
		probabilities. Lifespan can affect the short- and long-run behaviors of the 
		network, preventing formation if set too low, or inflating the size of the 
		SCC as it increases. We expect these steady-states can be analytically 
		approximated by functions of lifespan and binding probabilities, which is an 
		open avenue for future work.  Another investigation would be to what degree 
		lifespan is emphasized here as one of our minimal assumptions.  For example, 
		a species' death may continue to provide an affordance (easier access to 
		food, etc.) but these interactions would be lost in our model.  
		
		We can assume that evolution will favor a balanced $\rho$ ratio: too much 
		harm is 
		not sustainable. (This perspective is consistent with empirical 
		studies 
			of 
			Streptomyces communities in soil, where dense networks of antagonistic 
			and 
			growth-promoting interactions coexist, exhibiting balanced interaction
			probabilities, strong reciprocity, and rapid evolutionary turnover
			rather than
			static equilibrium~\citep{Vetsigian2011,mall2022limited}.) This balance
			point is not fixed: even within a single synthetic community, the ratio
			of inhibitory to facilitative interactions shifts with resource
			availability, becoming more antagonistic under nutrient limitation and
			more cooperative when resources are abundant~\citep{qiao2024nutrient},
			suggesting $\rho$ should be read as environmentally responsive rather
			than a fixed species trait. However, only help might
			hinder
		adaptability and evolvability ~\citep{GershensonEtAl2006}. To explore this
		hypothesis
		systematically, we plan to introduce evolutionary feedback explicitly in the 
		next 
		iteration of our model. Edge weights will dynamically evolve, strengthening 
		or 
		weakening interactions probabilistically over time, reflecting real-world 
		processes 
		of mutual adaptation and co-evolution. Preliminary results and theory 
		suggest this 
		enhancement should lead to rich dynamics, potentially revealing other novel 
		transitions or critical thresholds that influence community resilience and 
		connection.
		
		Finally, empirical validation remains paramount. A synthetic community is a
		defined set of isolates, assembled from a known species list and grown under
		controlled conditions, so that its membership is set by the experimenter
		rather than inferred after the fact. That control is what allows causal
		claims about interaction structure, which is why the approach has become the
		standard vehicle for testing community-assembly
		theory~\citep{vorholt2017establishing}. Reproducible, tunable synthetic soil
		communities are now an established experimental paradigm, from small defined
		consortia~\citep{coker2022reproducible} to high-throughput screens of
		hundreds of soil-derived communities~\citep{madsen2025synthetic}, and the
		approach has more than a decade of development behind it, with best practice
		across host and environmental systems now
		codified~\citep{grosskopf2014synthetic,mehlferber2024cross}. Its bottom-up
		strand asks which conditions are necessary to generate particular
		interaction patterns, and how higher-order community structure emerges from
		them, which is the question the present model poses \emph{in silico}.

		The natural test of the present model is a panel of synthetic communities
		whose members are chosen so that the
		fraction of antagonistic pairwise interactions differs from one community to
		the next, tracking in each whether a large, persistent cooperative core
		emerges and how its composition turns over. Droplet platforms that assay
		thousands of community combinations in parallel put a panel of that size
		within reach~\citep{kehe2019massively}. The model predicts a boundary in
		that panel: below it cores that keep recruiting, above it cores that hold a
		bounded size while their membership turns over. Will such networks look the
		same across different soil sources or starting conditions? Whatever daylight
		exists between our predictions and such empirical results would motivate, or
		afford, more interesting explorations.

		\subsection*{Acknowledgements}
		We thank Andrea Roli and Jérome Michaud for their insightful comments and 
		valuable 
		feedback, which greatly improved the manuscript.  We also thank Jan
		Dijksterhuis and Madalina-Maria Vita for discussions of the microbial
		communities that motivated this work.

		\subsection*{Code and data availability}
		The complete simulation code used to generate all results reported here is
		openly available at \url{https://github.com/maleygoes/life-finds-a-way},
		together with the parameter settings and instructions needed to reproduce
		the figures.  The model is implemented in Python using the NetworkX
		library.  This study is entirely computational and uses no empirical
		datasets.  A time-lapse animation of the age--degree relationship shown in
		Figure~\ref{scatter1} is provided as Supplementary Video~S1.

		\bibliographystyle{elsarticle-num-names}
		\bibliography{workslfawr3}
		
	\end{document}